\documentclass[12pt,a4paper]{article}
\input epsf
\usepackage{epsfig,
}
\def\ie{{i.e.}}
\def\IZ{\relax\ifmmode\mathchoice
{\hbox{\cmss Z\kern-.4em Z}}{\hbox{\cmss Z\kern-.4em Z}}
{\lower.9pt\hbox{\cmsss Z\kern-.4em Z}} {\lower1.2pt\hbox{\cmsss
Z\kern-.4em Z}}\else{\cmss Z\kern-.4em Z}\fi}
\def\IR{\relax{\rm I\kern-.18em R}}

\def\one{{\hbox{ 1\kern-.8mm l}}}

\newlength{\bredde}
\def\slash#1{\settowidth{\bredde}{$#1$}\ifmmode\,\raisebox{.15ex}{/}
\hspace*{-\bredde} #1\else$\,\raisebox{.15ex}{/}\hspace*{-\bredde}
#1$\fi}

\newsavebox{\zzzbar}
\sbox{\zzzbar}
  {\setlength{\unitlength}{0.9em}
  \begin{picture}(0.6,0.7)
  \thinlines
  \put(0,0){\line(1,0){0.6}}
  \put(0,0.75){\line(1,0){0.575}}
  \multiput(0,0)(0.0125,0.025){30}{\rule{0.3pt}{0.3pt}}
  \multiput(0.2,0)(0.0125,0.025){30}{\rule{0.3pt}{0.3pt}}
  \put(0,0.75){\line(0,-1){0.15}}
  \put(0.015,0.75){\line(0,-1){0.1}}
  \put(0.03,0.75){\line(0,-1){0.075}}
  \put(0.045,0.75){\line(0,-1){0.05}}
  \put(0.05,0.75){\line(0,-1){0.025}}
  \put(0.6,0){\line(0,1){0.15}}
  \put(0.585,0){\line(0,1){0.1}}
  \put(0.57,0){\line(0,1){0.075}}
  \put(0.555,0){\line(0,1){0.05}}
  \put(0.55,0){\line(0,1){0.025}}
  \end{picture}}

\newcommand{\ena}{\end{eqnarray}}
\newcommand{\beqa}{\begin{eqnarray}}
\newcommand{\eeqa}{\end{eqnarray}}

\newcommand{\half}{\frac{1}{2}}
\newcommand{\eq}[1]{(\ref{#1})}

\newcommand{\be}{\begin{equation}}
\newcommand{\ee}{\end{equation}}

\renewcommand{\b}{\beta}

\def\cs{{\cal S}}

\def\a{\alpha}
\def\b{\beta}

\def\d{\delta}
\def\e{\epsilon}
\def\f{\phi}

\def\k{\kappa}

\def\m{\mu}
\def\n{\nu}

\def\p{\pi}

\def\r{\rho}
\def\s{\sigma}
\def\t{\tau}

\def\L{\Lambda}

\begin{document}

\begin{titlepage}
\begin{flushright}
ITFA-2005-06 \\
hep-th/0503038
\end{flushright}
\vfill
\begin{center}
{\LARGE\bf Comments on BRST quantization of strings}
\\ \vskip 10.mm {  Ben Craps
and Kostas Skenderis } \\ \vskip 7mm {Instituut voor Theoretische
Fysica, Universiteit van Amsterdam,
Valckenierstraat 65,
1018 XE Amsterdam,
The Netherlands
}
\end{center}
\vfill

\begin{center}
{\bf ABSTRACT}
\end{center}
\vskip 7mm
The BRST quantization of strings is revisited and the derivation
of the path integral measure for scattering amplitudes is streamlined.
Gauge invariances due to zero modes in the ghost sector are taken into
account by using the Batalin-Vilkovisky formalism. This involves promoting
the moduli of Riemann surfaces to quantum mechanical variables on
which BRST transformations act. The familiar ghost and antighost zero mode
insertions are recovered upon integrating out auxiliary fields. In
contrast to the usual treatment, the gauge-fixed action including all zero
mode insertions is BRST invariant. Possible anomalous contributions
to BRST Ward identities due to
boundaries of moduli space are reproduced in a novel way. Two models are
discussed explicitly: bosonic string theory and topological gravity
coupled to the topological A-model.

\vfill
\hrule width 5.cm
\vskip 2.mm
{\small
\noindent  {\tt bcraps,skenderi@science.uva.nl}}
\end{titlepage}
\section{Introduction}

The BRST quantization of strings is a well studied subject, see 
\cite{Polchinski:1998rq,Deligne:1999qp,D'Hoker:1988ta} for textbook 
expositions and reviews.
In most works however global issues are either ignored 
or dealt with afterwards. For instance, a naive application 
of BRST quantization leads to amplitudes that all vanish due to 
ghost zero modes. To deal with these zero modes, one inserts 
a number of ghost fields in the path integral measure.
Historically, these insertions were first derived by 
a careful Fadeev-Popov 
analysis \cite{Friedan:1982is,Alvarez:1982zi,Moore:1985ix,D'Hoker:1985pj,Friedan:1985ge}. One could then show that 
the path integral in the presence of insertions is BRST invariant up to 
total derivatives in moduli space. This may look satisfactory,
but one should contrast this situation with the BRST quantization 
of quantum field theories, where the BRST method leads to an action and measure 
that are both BRST invariant (in the absence of BRST anomalies). 

We will show in this paper that one can incorporate global 
issues in the BRST quantization, leading to a straightforward 
derivation of the ghost insertions in the path integral. The main 
observation is that the existence of ghost zero modes implies 
that the gauge fixed action has additional gauge invariances, namely 
it is invariant under a variation of the ghosts proportional to their 
zero modes. The proper way to quantize the theory in such circumstances
is to use the BV or antifield formalism \cite{Batalin:1984jr,Henneaux:1992ig}. 
Essentially one 
introduces a new set of fields and following a well-established 
procedure arrives at the gauge fixed action.  

In the case at hand,
we will find that the new fields include
the moduli. Furthermore, upon integrating out a set of 
auxiliary fields one arrives at the usual ghost insertions 
in the path integral. In other words, the BRST-BV quantization 
automatically leads to an integral over the moduli space
with the correct measure. The resulting gauge-fixed action
(which incorporates all ghost insertions)
is by construction BRST invariant, so this treatment is 
exactly analogous to the QFT treatment. One should note
that the BRST-BV transformations differ from the ones 
appearing in most of the literature in that they also act on 
moduli (by a shift). The treatment of the moduli as quantum 
mechanical degrees of freedom on which BRST transformations act appeared before
in, for example, 
\cite{Baulieu:1987tq,Labastida:1988zb,Mansfield:1993qd,Becchi}. 
Our point of view is that the standard rules of quantization
require such a treatment and this leads to an automatic and simple
derivation of the path integral measure.

The current treatment also simplifies and streamlines 
the derivation of BRST Ward identities \cite{Friedan:1985ge, Cohen:1986mx,
Mansfield:1986it}. 
Since our formulation exactly 
parallels the BRST quantization of quantum field theories, one may 
just borrow the derivation of Ward identities in QFT where the identities
are simply derived by a change of variables in the path integral that 
amounts to a BRST shift, as in \cite{Mansfield:1993qd}. 
In the usual treatment one finds that the BRST trivial states decouple 
only if there is no contribution from the boundary of moduli space. This is
so because (in the usual treatment) the path integral measure is 
BRST invariant only up to total derivatives. In our case, we find the
same result but the derivation is somewhat different. 
In our discussion, the action and measure are BRST invariant.
However, BRST transformations shift the boundaries of the integral over moduli;
 in the  derivation of the Ward identity (which involves 
shifts of fields by their BRST variation) this leads to potential 
contributions from boundaries of moduli space (which may be ``at infinity''). 

These contributions from boundaries of the integration domain of some of the ``fields''
in the path integral is the main difference with 
usual quantum field theories. While in quantum field theory the only source of 
BRST anomalies is the Jacobian of the BRST transformations, in 
string theory anomalies may also originate from boundary terms. Actually
the discussion of anomalies in string theory usually involves the 
discussion of the boundary terms rather than the computation of Jacobians,
see for example \cite{Callan:1987px,Polchinski:1987tu} for discussions
of type I gauge anomalies.

Our considerations apply in general, but for concreteness we
present our discussion by means of two examples:
the bosonic string and topological strings. The emphasis in both cases 
is in new features and the derivation of the measure. The latter model
illustrates the issue of boundary contributions to Ward identities:
the model exhibits the so-called holomorphic anomaly. We will see
that the anomaly implies that the theory is gauge dependent.

This paper is organized as follow. In  section~2, we briefly 
review the basics of the BV formalism. In sections~3 and~4, we discuss bosonic 
and topological strings, respectively. Section 5 contains our conclusions.


\section{BV quantization}
In this section we review the basics of Batalin-Vilkovisky
(BV) or antifield quantization \cite{Batalin:1984jr,Henneaux:1992ig}.
In the BV formalism one constructs a BV action from which
both the gauge-fixed action and the BRST transformations can be
obtained.

The first step is the introduction of an appropriate
number of ghosts: for each local symmetry of the action
one introduces a set of ghost fields.
Let $\phi^i$ be the fields that a gauge invariant
classical action $\cs$ depends on. The gauge transformations
are given by
\be
\d \f^i = R^i_{\a_0} \e^{\a_0},
\ee
where $ \e^{\a_0}$ is the parameter of the transformation and
we use De Witt's condensed notation, \ie\ the indices can be discrete
or continuous; a repeated continuous index includes an integration.
Corresponding to these transformations, one introduces ghost
fields $C_0^{\a_0}$. If $R^i_{\a_0}$ are linearly independent in
a neighborhood of a stationary point $\f_0^i$ of the action $\cs$, then
the theory is ``irreducible'' and  $C_0^{\a_0}$ are all the
ghosts we need (apart from possible extraghosts in the non-minimal
sector, which we will discuss later). This will be the case for
the systems discussed in this paper.
If $R^i_{\a_0}$ is not of maximal
rank, \ie\ if there are non-zero solutions of
$R^i_{\a_0} Z^{\a_0}_{1 \a_1}|_{\f_0}=0$
($A|_{\f_0}$ denotes that $A$ is evaluated at the stationary point
$\f_0$ of the action $\cs$), the theory is ``reducible''
and one needs an  additional set of ghosts. Specifically,
reducibility implies that the action for the ghosts
$C_0^{\a_0}$ has a new gauge invariance given by
$\d C_0^{\a_0} = Z^{\a_0}_{1 \a_1} \e_1^{\a_1}$. This leads to the
introduction of ghosts-for-ghosts $C_1^{\a_1}$. Similarly,
if $Z^{\a_0}_{1 \a_1}$ is not of maximal rank the action
for the ghosts-for-ghosts will have a gauge invariance and
we need a new set of ghosts $C_2^{\a_2}$, and so on.

Let $\f^A$ denote the collection of the fields $\f^i$ and
of all ghosts $C_0^{\a_0}, C_1^{\a_1}, ... $. We introduce
an antifield $\f_A^*$ for each field $\f^A$. The antifield $\f_A^*$ has
opposite statistics compared to the corresponding field $\f^A$,
and its ghost number is
\be
{\rm gh}(\f_A^*)=-{\rm gh}(\f^A)-1.
\ee
We then define an odd graded Lie bracket,
the antibracket:
\be
(A,B) = \frac{\delta^R A}{\delta \f^A} \frac{\delta^L B}{\delta \f_A^*}
- \frac{\delta^R A}{\delta \f_A^*} \frac{\delta^L B}{\delta \f^A},
\ee
where right and left derivatives are defined by
$
\delta F = (\delta^R F/\delta z) \delta z = \delta z (\delta^L F/\delta z).
$
The minimal action $S_{min}[\f^A,\f_A^*]$ is given by the solution to
the master equation,
\be
(S,S)=0,
\ee
with the boundary condition that $S[\f^A,0]=\cs[\f^i]$. If the gauge
algebra closes off-shell (as in our examples) $S$ is linear in the antifields.

To gauge fix the theory, we need to introduce additional
fields, the so-called non-minimal sector:
for each set of ghosts $C_i^{\a_i}$ we introduce a set
of antighosts $b_i^{\b_i}$,\footnote{In general, the ghost field
transforms in the adjoint representation of the gauge algebra and 
the antighost in the co-adjoint. For finite dimensional Lie algebras 
corresponding to 
compact Lie groups the adjoint and co-adjoint representations 
are the same. In general, however, the indices $\b_i$ are
different from the indices $\a_i$.} a set of auxiliary fields
$\pi_i^{\b_i}$ and corresponding antifields. Finally, if the
gauge fixing condition is such that the ghost action
has a new gauge invariance with the antighosts being the gauge
fields, extraghosts, corresponding
auxiliary fields and
antifields are required. We will not review the general case
here; the interested reader may consult the original literature and the
reviews. If the system is irreducible, 
as are the systems discussed in this paper, one only needs antighosts
$b_0^{\beta_0}$, auxiliary fields $\pi_0^{\beta_0}$, extraghosts $C_0^k{}'$,
auxiliary fields $\pi_0^k{}'$, and antifields for all these fields.

The minimal solution to the master equation is now extended to incorporate the
non-minimal sector,
\be
S=S_{min} + b^{0*}_{\b_0} \pi_0^{\b_0} +  C^{*}_{0k}{}'
\pi_0^k{}',
\ee
where we only included a single set of antighosts and extraghosts.
Gauge fixing is achieved by first performing a canonical transformation,
\ie\ a transformation $\f^A \to \f^A{}'(\f^B,\f_B^*),
\f^*_A \to \f^*_A{}'(\f^B,\f_B^*)$
that preserves the antibracket, and then setting the antifields to zero, as
we now explain.

Canonical transformations are
always generated by a fermionic generator $\Psi$, the so-called
gauge fixing fermion,
\be
\f^A{}'=e^\Psi \f^A \equiv \f^A + (\Psi, \f^A) + \frac{1}{2}
(\Psi, (\Psi, \f^A)) + \cdots
\ee
and similarly for $\f^*_A{}'$. We define the BV action
\be
S_{BV}[\f^A, \f_A^*] = S[\f^A{}', \f_A^*{}'].
\ee
If, as will be the case in this paper, $\Psi$ depends only on the
fields, not on the antifields, we obtain
\be
S_{BV}[\f^A, \f_A^*] = S[\f^A, \f_A^*
+ \frac{\partial \Psi}{\partial \f^A}].
\ee
This action is invariant under the following BRST transformation
acting on both the fields and the antifields,
\be \label{brst}
\d_{BRST} \f^A = (\f^A, S_{BV}\L), \qquad
\d_{BRST} \f^*_A = (\f^*_A, S_{BV}\L),
\ee
where we introduced a constant anticommuting variable $\L$ such that
$\d_{BRST}$ is a derivation rather than an antiderivation.
This transformation is nilpotent off-shell.

The gauge-fixed action is obtained from the BV action
by simply setting the antifields to zero,
\be \label{gf}
S_{gf}[\f^A] = S_{BV}[\f^A,0].
\ee
This action is invariant under the BRST transformations
in (\ref{brst}) with the antifields set to zero,
$\d_{BRST} \f^A = (\f^A, S_{BV}\L)|_{\f_A^*=0}$.

We now specialize to irreducible theories, the case of interest in this paper.
The gauge fixing fermion is usually taken to be of the form
\be \label{psi}
\Psi=b_0^{\b_0} \chi_{\b_0}(\f^{i}) + b_0^{\b_0} \s_{\b_0 k'}
C_0^k{}'.
\ee
In the absence of the last term this gauge fixing fermion will
lead to a $\d$-function gauge fixing that sets $\chi_{\b_0}(\f^i)=0$.
The term with the extraghost is necessary if the
gauge fixing condition leads to a ghost action with a
new gauge invariance acting on the antighost, as we now
explain. Following the steps we outlined above and ignoring
for the moment the last term in $\Psi$
one arrives at the ghost action
\be
b_0^{\b_0} \frac{\partial \chi_{\b_0}}{\partial \f^i} R^i_{\a_0}
C^{\a_0}.
\ee
If the matrix $A_{\b_0 \a_0} = \partial_{\f^i} \chi_{\b_0} R^i_{\a_0}$
has a left zero eigenvalues, $\bar{Z}^{\b_0}_k A_{\b_0 \a_0}{=}0$, then the
ghost action is invariant under the symmetry
$\d b_0^{\b_0} = \bar{Z}^{\b_0}_k \e_0^k{}'$. The extraghost
$C_0^k{}'$ is introduced in order to deal with this gauge
invariance, and the last term in (\ref{psi}) is the
corresponding gauge fixing condition. The matrix $\s_{\b_0 k'}$
is any convenient matrix of maximal rank.

Notice that in many cases studied in the literature the number of
right zero eigenvalues of $R^i_{\a_0}$ is the same as the number of 
left zero eigenvalues of $A_{\b_0 \a_0} 
=\partial_{\f^i} \chi_{\b_0} R^i_{\a_0}$ and as result
extra-ghosts and ghosts-for-ghosts appear simultaneously.
In general, however, the two need not coincide and this is what 
happens in the case of interest to us. In such cases one can have 
extra-ghosts without ghosts-for-ghosts (or vice versa).

\section{BRST symmetry of the bosonic string}
\subsection{Gauge invariant action}\label{sub:inv}
We now apply the BV formalism to the sigma model describing bosonic
closed string theory.
The fields $\phi^i$ include the worldsheet metric $g_{ab}(\sigma)$ and
the scalar fields $X^\mu(\sigma)$ corresponding to spacetime coordinates.
If the spacetime is flat, the worldsheet action is \cite{Polchinski:1998rq}
\be\label{WSactionnoVO}
{\cal S}[g_{ab},X^\m]={1\over4\pi\alpha'}\int_M d^2\sigma
g^{1/2}g^{ab}\partial _a X^\mu\partial _bX_\mu\,+{\lambda\over4\pi} \int_M
d^2\sigma g^{1/2}R,
\ee
where $M$ is a Riemann surface.

The gauge symmetry of the action \eq{WSactionnoVO} consists of Weyl
rescalings of the worldsheet metric and diffeomorphisms of the
worldsheet. The infinitesimal gauge variation is given by
\be\label{gaugevar}
\delta g_{ab}=2 \omega
g_{ab}+\nabla_a \e_b+\nabla_b\e_a, \qquad
\delta X^\m = \e^a \partial_a X^\mu,
\ee
where $\omega(\sigma)$ and $\e^a(\sigma)$
parametrize the infinitesimal Weyl transformation and
diffeomorphism, respectively (they were denoted $\e^{\a_0}$ in the
previous section).

In fact, we will really be interested in computing correlation
functions of vertex
operators, corresponding to string scattering amplitudes.
We proceed by introducing sources $\r^i$ with Weyl weight one 
that couple to the
vertex operators $V_i$, which are scalar functionals with 
Weyl weight minus one. 
The worldsheet action is then modified as follows,
\be
\label{WSaction}
S_0[g_{ab},X^\m,\s_i^a;\r^i]=
{\cal S}+\sum_{i=1}^n \r^i
V_i(\sigma_i).
\ee
This way, \eq{WSaction} is invariant
under diffeomorphisms and Weyl transformations if we accompany
the usual action of those
transformations with an explicit shift of $\sigma_i$,
\be
\d \s_i^a = - \e^a(\s_i),
\ee
and we take the sources to be invariant under diffeomorphisms.
Differentiating with respect to the sources leads to an insertion of
the vertex operators in the path integral.\footnote{
Note that for an $n$-point function we introduce $n$ sources
even when some of the operators are the same
and the sources are always treated infinitesimally, i.e.\
we only differentiate once with respect to each source and then
set all sources to zero.} One of our tasks below will be
to show that this insertion is accompanied by either a ghost
insertion or an integration over $\sigma_i$.

In \eq{WSaction} we consider the $\sigma_i^a$ to be fields on
the same footing as the fields $g_{ab}$ and $X^\mu$, except that $\sigma_i^a$
are constant fields, i.e.\ they do not depend on the coordinates $\sigma^a$. In other
words, in the path integral we integrate over $\s_i^a, g_{ab}, X^\m$
(and ghosts, antighosts and auxiliary fields introduced
during the gauge fixing procedure, as we explain below).
This may seem unusual, but we will see that it leads to an elegant
derivation of the gauge-fixed path integral.

The role of the index $i$ of the fields $\phi^i$ in the previous section
(not to be confused with the index $i$ used in the present section) is now
played by $((ab),\sigma)$ (the indices and argument of $g_{ab}(\sigma)$),
$(a,i)$ (the indices of $\sigma_i^a$) and $(\mu,\sigma)$ (the index and argument
of $X^\mu(\sigma)$).
Similarly, the index $\alpha_0$ in the previous section is now
replaced by $\sigma'$ (the argument of $\omega(\sigma')$)
and $(c,\sigma')$ (the index and argument of
$\e^c(\sigma')$). The matrix $R$ is then given by
\beqa
R_{\sigma'}^{(ab),\sigma}&=&2g_{ab}\delta(\sigma-\sigma'),\\
R_{c,\sigma'}^{(ab),\sigma}&=&(\nabla_a\delta_b^c+\nabla_b\delta_a^c)
\delta(\sigma-\sigma'),\\
R_{\sigma'}^{a,i}&=&0,\\
R_{c,\sigma'}^{a,i}&=&-\delta_c^a \delta(\sigma_i-\sigma'),\\
R_{\sigma'}^{\mu,\sigma}&=&0,\\
R_{c,\sigma'}^{\mu,\sigma}&=&\partial_cX^\mu\delta(\sigma-\sigma').
\eeqa

\subsection{Gauge fixed action}

The first step in the BV procedure is to introduce ghost and auxiliary fields.
In our case we have fermionic ghost fields $C_\omega(\sigma)$ (for the Weyl
transformations) and $c^a(\sigma)$ (for the diffeomorphisms);
these ghost fields were denoted $C_0^{\alpha_0}$ in the previous
section. We also introduce several pairs of auxiliary fields.
A first pair corresponds to $\tilde b^{ab}(\sigma)$ (a fermionic antighost) and
$\pi^{ab}(\sigma)$ (a boson with ghost number zero). A second pair is
formed by constant fermionic antighosts $b^j_a$ and ghost number zero
bosons $p_j^a$ for a set of values
\be
(a,j)\in f,
\ee 
where $f$ will correspond to the set of fixed vertex operator coordinates.
In particular, $(a,j)=1,\ldots,\k$ and $\k$ is the number of
conformal Killing vectors: $\k=6$
for a Riemann surface of genus zero, $\k=2$ for genus one and $\k=0$
for higher genus.
A third pair consists of constant extraghosts $\tau^k$
(bosons of ghost number
zero) and corresponding fermionic fields $\xi^k$ of ghost number
one. Here $k=1,\ldots,\mu$, with $\mu$ the number of metric
moduli:
$\mu=0$ for genus zero, $\mu=2$ for genus one and $\mu=6g-6$ for
higher genus. In fact, the extraghosts $\tau^k$ will be
interpreted as metric moduli.

In the antifield formalism, the action takes the form
\beqa
S&=&S_0+\int d^2\sigma\, \left( g_*^{ab}
\left[2C_\omega(\sigma) g_{ab}(\sigma)
+\nabla_ac_b(\sigma)+\nabla_bc_a(\sigma)\right]
+ \tilde b^*_{ab}(\sigma)
\pi^{ab}(\sigma)\ \right. \nonumber \\
&& \left. \ \ \ \ \qquad \qquad -c^*_a(\s) c^b \partial_b c^a(\s)
-C^*_w(\s) c^b \partial_b C_w(\s) + X_\m^* (\s) c^a \partial_a X^\m
\right) \nonumber\\
&&\qquad - \sum_{i=1}^{n} \sigma_a^{i*}c^a(\sigma_i)
+ \sum_{(a,j)\in f} b_a^{j*} p^a_j
+ \sum_{k=1}^\mu\tau_k^*\xi^k ,\label{actionanti}
\eeqa
where the fields in the last line are constant fields. (The summation
over the repeated vector indices is implicit throughout this paper.)
Notice
that the antifields in (\ref{actionanti}) transform as densities.
One could have introduced instead explicit factors of $g^{1/2}$, but
the present convention simplifies some of the computations
below.

We now discuss gauge fixing. 
Equivalence classes of metrics under diffeomorphisms and Weyl transformations 
are labelled by coordinates $\tau^k$. We choose a smooth set of 
reference metrics
$\hat g_{ab}(\tau^k;\s)$ (which can be and are chosen to have constant 
curvature).
Further, we choose a collection $\hat\sigma_i^a$ of reference values for
those fields $\sigma_i^a$ with $(a,i)\in f$.
We can then gauge fix the action as follows.
For the gauge fermion we make the following choice:
\be \label{gfixing}
\Psi=
{1\over4\pi}\int d^2\sigma\, \tilde b^{ab}(\sigma)(g_{ab}(\sigma)
-\hat g_{ab}(\tau^k;\sigma))
+\sum_{(a,j)\in f} b_a^j(\sigma^a_j-\hat\sigma_j^a).
\ee
The first term imposes the gauge condition
\be
\chi_{ab}(\sigma)=g_{ab}(\sigma)-\hat g_{ab}(\tau^k;\sigma)=0.
\ee
Let us now motivate why we introduced the moduli $\tau^k$ as
(constant) fields in the action. 
If we had not done so, but merely
considered $\tau^k$ as parameters instead of fields,
the gauge fixing fermion (\ref{gfixing}) would have led (following the steps we
discuss below) to the usual ghost action\footnote{Note that $\tilde{b}^{ab}$
transforms as a density (in addition to the transformation implied by its
indices). In the literature one often uses a tensor field $b_{ab}$ related
to our $\tilde{b}^{ab}$ by $\tilde{b}^{ab}= \sqrt{g} g^{ac} g^{bd} b_{cd}$.}
\be
S_{gh} = \frac{1}{2\pi}\int d^2 \s \tilde{b}^{ab} (P_1 c)_{ab},
\ee
where the operator $P_1$ maps vectors to symmetric
traceless tensors,
$(P_1c)_{ab}=\half(\nabla_a c_b+\nabla_b c_a
-g_{ab}\nabla_c c^c)$. This ghost action has
additional invariances because of the zero modes
of $P_1$ and $P_1^\dagger$: the action is
invariant under a shift of $c^a$ by a conformal Killing
vector as well as under a shift of $\tilde{b}^{ab}$ by a holomorphic
quadratic differential. The symmetry due to ghost zero modes is gauge fixed by fixing the
positions of $\k$ vertex operators. This is
the origin of the last term in (\ref{gfixing})
which enforces the gauge condition,
\be
\chi_j^a = \s_j^a - \hat{\s}_j^a=0, \ \ (a,j)\in f.
\ee
In the absence of vertex operators in (\ref{WSaction}),
the presence of ghost zero modes would imply that the
gauge algebra is reducible and one would proceed by
introducing ghosts-for-ghosts, as described in the
previous section.
The antighost zero modes also lead
to an invariance of the action, as discussed below
(\ref{psi}). To deal with this
invariance we interpret the moduli as extraghost fields, playing
the role of ${C_0^k}'$ in (\ref{psi}).
 Compared
with the previous section, $\partial_k \hat{g}_{ab}(\tau^k;\s)$
is what we called $\s_{\b_0 k'}$. Recall that the tangent
space to the moduli space at $\hat{g}_{ab}(\tau^k;\s)$ (which
is a metric of constant curvature) is spanned by the
quadratic differentials ${\rm Ker} P_1^\dagger$, so $\partial_k\hat{g}_{ab}(\tau^k;\s)$
has indeed maximal rank.

The gauge fixing fermion in (\ref{gfixing}) leads
to the gauge fixed action
\beqa \label{actionSigma}
S_{gf}&=&S_0\\
&&+{1\over4\pi}\int d^2\sigma\, \tilde b^{ab}(\sigma)
\left[2C_\omega(\sigma) g_{ab}
(\sigma) +\nabla_ac_b(\sigma)+\nabla_bc_a(\sigma)-\xi^k\partial_k\hat g_{ab}(\tau;\sigma)\right]\nonumber \\
&&+{1\over4\pi}\int d^2\sigma\, \pi^{ab}(\sigma)
\left[g_{ab}(\sigma)-\hat g_{ab}(t_0;\sigma)\right]\nonumber\\
&&+\sum_{(a,j)\in f}\left[- b_a^j c^a(\sigma_j)
+p_a^j(\sigma^a_j-\hat\sigma_j^a)\right]. \nonumber
\eeqa
This concludes the construction of the gauged fixed action. The gauged fixed
action one often finds in the literature does not contain the last line
and the last term in the first line. In addition, $C_\omega$ and $\p^{ab}$
have usually been integrated out.

We thus arrive at the following generating functional of string amplitudes,
\be
Z[\r^i]=\int d \mu e^{-S_{gf}}
\ee
where
\be \label{measure}
d \mu = \prod_{i=1}^n d^2\s_i \sqrt{g(\s_i)} \prod_{j=1}^\k (d b^j d p_j)
\prod_{k=1}^\mu (d\tau^k d \xi^k) [d X^\mu] [d g_{ab}] [d \pi_{ab}]
[d \tilde{b}^{ab}] [d C_w] [d c^a].
\ee
Many of the fields are auxiliary and can be integrated out, as we discuss
in subsection~\ref{sub:aux}.

\subsection{BRST transformations}

The action \eq{actionSigma} is designed to satisfy the master equation and
is thus BRST invariant by construction. One can verify this almost
by inspection as we now discuss. Recall that the  BRST transformation of
a field $\phi^A$ appearing in the action can be read off from the BV action:
\be
\delta_{BRST}\phi^A={\d^L  S\over\d  \phi_A^*}\Lambda,
\ee
where $\Lambda$ is an anticommuting parameter.
In particular,
\beqa
\d_{BRST} X&=&c^a\partial_aX\Lambda, \label{modvar} \\
\d_{BRST} g_{ab}&=&(c^c\partial_cg_{ab}+\partial_ac^cg_{cb}+\partial_bc^cg_{ac}
+2C_\omega g_{ab})\Lambda,\nonumber  \\
\d_{BRST} c^b&=&-c^a\partial_ac^b\Lambda,\nonumber \\
\d_{BRST} C_\omega&=&-c^a\partial_aC_\omega\Lambda,\nonumber \\
\delta_{BRST}\tau^k&=&\xi^k\Lambda,\nonumber \\
\delta_{BRST}\tilde b^{ab}(\sigma)&=&\pi^{ab}(\sigma)\Lambda,\nonumber \\
\d_{BRST} \s_i^a &=& - c^a(\s_i) \L,\nonumber \\
\delta_{BRST}b_a^j&=&p_a^j\Lambda, \nonumber 
\eeqa
and $\pi^{ab}, \xi^k, p_a^j, c^a(\s^i)$ are BRST invariant\footnote{
Nilpotency of the BRST transformation requires
$\d_{BRST} c^a(\s_i)=0$, and one can verify that this holds,
$\d_{BRST} c^a(\sigma_i)=
-c^b\partial_bc^a(\sigma_i)\Lambda+\partial_bc^a(\sigma_i)
[-c^b(\sigma_i)\Lambda]=0$.}.

Defining the BRST charge $Q_B$ by
\be
\delta_{BRST}\Phi=\{Q_B,\Phi\}\Lambda,
\ee
the gauge fixed action (\ref{actionSigma}) can be written as
\be\label{gfaction}
S_{gf}=S_0+\left\{
Q_B,{1\over4\pi}\int d^2\sigma\,b^{ab}[g_{ab}-\hat g_{ab}(\tau)]
+\sum b_a^j(\sigma^a_j-\hat \sigma^a_j)
\right\}.
\ee
This equation is familiar in BRST quantization: to gauge fix, one
adds a BRST exact term to the original action. The reason we went
through the more elaborate BV formalism is to motivate the fact
that that the moduli $\tau^k$ transform under the BRST symmetry.
{}From (\ref{gfaction}) and the nilpotency of $\delta_{BRST}$, it is
clear that the full gauge fixed action is invariant under our BRST
transformations.

\subsection{Integrating out auxiliary fields}\label{sub:aux}

To make contact with the textbook expressions for the gauge fixed action
and of string amplitudes, we now integrate out a number of
auxiliary fields from the action (\ref{actionSigma}).

We will integrate out the fields $C_\omega$, 
$\tilde b^a_a=g_{ab}\tilde b^{ab}$, $b^i_a$,
$\pi^{ab}$, $p^i_a$, $g_{ab}$, $\xi^k$, $\xi_i^a$ and $\rho^a$.
Performing the functional over the fermionic field $C_\omega$ produces insertions of the
various modes of $\tilde b^a_a(\sigma)$ in the path integral. These insertions effectively set
$\tilde b^a_a$ equal to zero in the action; they disappear upon performing the functional integral
over $\tilde b^a_a(\sigma)$.
As mentioned before, integrating out
the bosonic field $\pi^{ab}(\sigma)$ enforces
\be\label{fixedmetrictris}
g_{ab}=\hat g_{ab}(\tau^k).
\ee
Similarly, integrating out the bosonic fields $p_a^i$ sets
\be
\sigma_i^a=\hat \sigma_i^a\ \ \ {\rm if}\ (a,i)\in f.
\ee
Finally, integrating out $\xi^k$, $p^a_i$ and $b^i_a$ and differentiating
with respect to $\rho_i$ and setting $\rho_i=0$,  leads to the
familiar path integral
\beqa
\langle V_1(k_1) \cdots V_n(k_n)\rangle
&=&\sum_{g=0}^\infty \int
d^\mu\tau\int[dX\,d\tilde b\,dc]\exp(-{\cal S}-S_{gh})\nonumber\\
&&\times\prod_{(a,i)\in f}
\sqrt{\hat g(\hat\sigma_i)} 
c^a(\hat{\sigma}_i) V_{i}(k_i,\hat{\sigma}_i) 
\prod_{k=1}^\mu {1\over4\pi}(\tilde b,\partial _k\hat g)\nonumber\\
&&\times \prod_{(a,i)\not\in f}\int {d\sigma}^a_i
\sqrt{\hat g({\sigma}_i)} V_{i}(k_i,\sigma_i),
\label{Polchtilde}
\eeqa
where 
$(\tilde b,\partial _k\hat g) = \int d^2 \s  \,
\tilde{b}^{ab} \partial _k\hat g_{ab}$.
The ghost and antighost insertions in the path integral are explained 
as follows. Starting from (\ref{actionSigma}) and 
integrating out $p_a^i$ and $b_a^i$ introduces $\k$ delta functions
that fix the positions of $\k$ vertex operators and inserts the ghost
$c^a(\hat\sigma_i)$ in the path integral. This is the familiar
ghost insertion that accompanies fixed vertex operators. The
antighost insertions come from the integral over $\xi^k$.

\subsection{Do BRST exact states decouple?}

Since the full gauge fixed action \eq{actionSigma} is BRST
invariant, one might think that if the path integral measure is
invariant, BRST exact states should decouple. The formal argument,
familiar from the derivation of BRST Ward identities in gauge theories,
goes as follows.
Let $V[\f^A]$ be an arbitrary function of all fields and consider
\be
\langle V[\f^A] \rangle_{\r^i} \equiv \int d \mu V[\f^A]
e^{-S_{gf}[\f^A;\r^i]},
\ee
where $\f^A$ denotes collectively all fields we integrate over in the
path integral (see (\ref{measure})) and $\langle V[\f^A] \rangle_{\r^i}$
indicate the 1-point function of $V[\f^A]$ in the present of sources,
i.e.\ this expression encompasses arbitrary $n$-point functions of 
$V[\f^A]$ with other vertex operators (obtained by 
differentiating the 1-point function with respect to the sources and then setting
the sources to zero). We now change variables
\be\label{changevar}
\phi^A{}' = \phi^A + \d_{BRST} \phi^A.
\ee
Provided that the measure is invariant, i.e.\ that there are no
BRST anomalies, one finds
\be\label{decoup}
\int d \mu' V[\f^A{}'] e^{-S_{gf}[\f^A{}';\r^i]}
=\int d \mu V[\f^A] e^{-S_{gf}[\f^A{};\r^i]}
+ \int d \mu (\d_{BRST} V[\f^A]) e^{-S_{gf}[\f^A{};\r^i]},
\ee
which implies
\be \label{WI}
\langle \delta_{BRST} V[\f^A] \rangle_{\r^i} =0.
\ee
In other words, BRST exact states seem to decouple from the correlation 
functions of arbitrary number of BRST invariant vertex operators.

However, there is a loophole in this argument. For the moduli
fields, the transformation \eq{changevar} reads
\be\label{tautransf}
\tau^k{}'=\tau^k+\xi^k\Lambda,
\ee
that is, the change of variables shifts the moduli. Now if the
integral over moduli gets contributions from boundaries of the
integration domain (which may be at infinity), i.e.\ from boundaries
of moduli space, the shift gives rise to extra boundary terms on
the right hand side of \eq{decoup}. So BRST exact states only
decouple if the boundary terms vanish.

To see this more explicitly, single out a modulus $\tau^0$ and assume
that the moduli space has a boundary at 
$\tau^0=\tau^{0(f)}$: $\tau^0\leq \tau^{0(f)}$. The boundary may
be at infinity or at finite value. (One may similarly incorporate 
a boundary located at the lower end of the integration domain of $\t^0$.)
Let us write the path integral
measure as $d\mu=d\tau^0d\tilde\mu$, 
and similarly $V[\phi^A]=V[\tau^0,\tilde\phi^A]$,
where a tilde denotes that $\tau^0$ is excluded. 
Running the previous argument, noting
that $\delta\tau^0=\xi^0\Lambda$ and keeping track of 
contributions from the boundary
of the $\tau^0$ integration domain, one finds
\beqa\label{decoupbndy}
&&\int^{\tau^{0(f)}}d\tau^0{}'\int d \tilde\mu' V[\tau^0{}',\tilde\f^A{}'] 
e^{-S_{gf}[\tau^0{}',\tilde\f^A{}';\r^i]}\\
&&=\int^{\tau^{0(f)}-\xi^0\Lambda}d\tau^0\int d \tilde\mu 
V[\tau^0+\xi^0\Lambda,\tilde\f^A+\delta\tilde\f^A] 
e^{-S_{gf}[\tau^0+\xi^0\Lambda,\tilde\f^A+\delta\tilde\f^A;\r^i]}\nonumber\\
&&=\int^{\tau^{0(f)}}d\tau^0\int d \tilde\mu V[\tau^0,\tilde\f^A] 
e^{-S_{gf}[\tau^0,\tilde\f^A;\r^i]}\nonumber\\
&&\ \  + \int^{\tau^{0(f)}}d\tau^0\int d \tilde\mu \left(\delta_{BRST}
V[\tau^0,\tilde\f^A]\right) 
e^{-S_{gf}[\tau^0,\tilde\f^A;\r^i]}\nonumber\\
&& \ \ 
-\int d \tilde\mu \xi^0\Lambda V[\tau^{0(f)},\tilde\f^A]
e^{-S_{gf}[\tau^{0(f)},\tilde\f^A;\r^i]}\nonumber
\eeqa
which implies
\be \label{nwi}
\langle \delta_{BRST} V[\f^A] \rangle_{\r^i} = 
\langle \partial_{\t^0} \left(\xi^0\Lambda V[\f^A]\right) \rangle_{\r^i}
\ee
Now  because of the $\xi^0$ 
insertion in the right hand side of (\ref{nwi}), the integral over
$\xi^0$ will not bring down the usual $(\tilde b,\partial_0\hat g)$ 
antighost insertion: the boundary term has one less antighost insertion 
compared to the bulk terms.

In the textbook treatment of BRST quantization of strings, these boundary
terms arise in a different way. There the BRST transformations do
not act on moduli, and it is the gauge fixed action excluding
terms giving rise to antighost insertions that is BRST invariant.
The antighost insertions themselves are not invariant because the
antighosts transform into the stress tensor. To see this, use \eq{modvar},
combined with the equation of motion for $g_{ab}$:
\be\label{Thetadef}
\pi^{ab}=-4\pi{\d S_2\over \d g_{ab}}=g^{1/2}\, \Theta^{ab},
\ee
where $\Theta^{ab}$ is the stress tensor of the action \eq{actionSigma} with
the term involving $\pi^{ab}$ omitted.
The resulting stress tensor insertion in turn gives rise to a
total derivative on moduli space, which upon integration over
moduli space leads to boundary terms.

\subsection{Summary}

In this section, we have used the Batalin-Vilkovisky formalism to
derive a manifestly BRST invariant action for the bosonic string.
This action includes terms that give rise to ghost and antighost
insertions upon integrating out auxiliary fields. In the path
integral, the integrals over moduli and vertex operator positions
are automatically present, because moduli and vertex operator
positions are considered to be (constant) fields in the action. A
notable feature of this formalism is that BRST transformations act
on moduli; this leads to potential non-decoupling of BRST exact
states due to contributions from boundaries of moduli space.


\section{Topological strings}

In this section we consider topological gravity coupled
to the topological sigma A-model. This model
exhibits the so-called holomorphic anomaly \cite{Bershadsky:1993cx}:
certain BRST exact terms do not decouple because of contributions from 
boundary terms. One of the motivations for this work was to 
understand the implications of the anomaly. Usually breaking of BRST invariance
in QFT implies lack of renormalizability and unitarity, but these
do not seem to be an issue for topological theories. Another 
implication of the non-decoupling
of  BRST exact states is that the quantum theory is gauge 
dependent: shifting the gauge fixing term by the BRST exact term 
corresponding to the state 
that does not decouple leads to an inequivalent theory. Thus, the holomorphic 
anomaly implies that topological string theory is gauge dependent.
It is unclear to us what is the proper worldsheet interpretation of this fact,
but we note that the holomorphic anomaly has also been linked to a
quantum version of background independence \cite{Witten:1993ed} (see also 
\cite{Dijkgraaf:2002ac}). 

\subsection{The topological sigma model}

In this subsection, we briefly review the A-model topological
sigma model \cite{Witten:1988xj} as constructed in
\cite{Baulieu:1989rs}. We restrict our attention
to target spaces that are Calabi-Yau manifolds.

The starting point is the action
\be\label{BSaction}
I[X]=\int_M (\omega_{\mu\nu}+iB_{\mu\nu})dX^\mu \wedge dX^\nu=\int_M dz d\bar z
(\omega_{\mu\nu}+iB_{\mu\nu})\partial X^\mu\bar\partial X^\nu,
\ee
where the worldsheet $M$ is a Riemann surface, 
$\omega$ the K\"ahler form of the
K\"ahler metric $G_{\mu\nu}$ on the target space $N$,
\be
\omega_{i\bar j}=-iG_{i\bar j},\ \ \ \omega_{\bar j i}=iG_{i\bar
j},
\ee
$B$ an antisymmetric tensor potential with zero field strength,
and $X$ a smooth map from $M$ to $N$. (The B-field was not present in 
\cite{Baulieu:1989rs}, but will play a role when we discuss the holomorphic anomaly.
Its inclusion complexifies the space of K\"ahler deformations of $N$.)
The action is independent of the worldsheet metric and only depends on the
cohomology class of the  K\"ahler form and the homotopy 
class of the map $X$. As a consequence, it has a
gauge symmetry corresponding to arbitrary small deformations
of $X$:
\be\label{gaugesymm}
\delta X^\mu=\epsilon^\mu.
\ee
This gives rise to the BRST symmetry
\beqa\label{BRSTBS}
\delta_S X^\mu&=&\psi^\mu,\\
\delta_S \psi^\mu&=&0,\nonumber\\
\delta_S \bar\psi^\mu&=&b^\mu,\nonumber\\
\delta_S b^\mu&=&0,\nonumber
\eeqa
where $\psi^\mu$ is a ghost field, $\bar\psi^\mu$ an antighost
field and $b^\mu$ an auxiliary field.

In terms of the complex structure $J$ of the target space (given by 
$J^\m{}_\n =G^{\mu\rho}\omega_{\rho\nu}$),  define
\be
\dot X=(1-iJ)\bar\partial X+(1+iJ)\partial X,
\ee
or in other words
\be
\dot X^i=2\bar\partial X^i,\ \ \ \dot X^{\bar j}=2\partial X^{\bar
j}.
\ee
In \cite{Baulieu:1989rs} the following gauge fixing was chosen:
\beqa
I_{gf}&=&\int dz d\bar z
(\omega_{\mu\nu}+iB_{\mu\nu})\partial X^\mu\bar\partial X^\nu\nonumber\\
&&-{i\over 2}\int dz d\bar z\,\delta_S \left\{
\bar\psi^\mu(G_{\mu\nu}\dot X^\nu-{1\over2}G_{\mu\nu}b^\nu
+{1\over2}\Gamma_{\mu\sigma\rho}\bar\psi^\sigma\psi^\rho)
\right\},
\label{BSactiongf}
\eeqa
where 
$
\Gamma_{\mu\sigma\rho}={1\over2}(\partial_\sigma G_{\mu\rho}
+\partial_\rho G_{\mu\sigma}-\partial_\mu G_{\sigma\rho})
$
is the Christoffel symbol. Using \eq{BRSTBS}, the gauge fixed action
can be written as
\beqa
I_{gf}&=&\int dz d\bar z
(\omega_{\mu\nu}+iB_{\mu\nu})\partial X^\mu\bar\partial X^\nu\nonumber\\
&&-{i\over 2}\int dz d\bar z\,\bigl\{
-\half G_{\mu\nu}b^\mu b^\nu+b^\mu(G_{\mu\nu}\dot X^\nu
+\Gamma_{\mu\sigma\rho}\bar\psi^\sigma\psi^\rho)\nonumber\\
&&\ \ \ \ \ \ \ \ \ \ -\bar\psi^\mu(G_{\mu\nu}\dot\psi^\nu+\partial_\rho G_{\mu\nu}\dot X^\nu\psi^\rho)
+\half\bar\psi^\mu\psi^\rho\bar\psi^\sigma\psi^\tau\partial_\tau\Gamma_{\mu\sigma\rho}
\bigr\}.
\eeqa
We now eliminate the auxiliary field $b^\mu$ using its equation of motion
\be
b^\mu=\dot X^\mu+\Gamma^\mu{}_{\sigma\rho}\bar\psi^\sigma\psi^\rho
\ee
and obtain
\beqa
I_{gf}&=&-i\int dz d\bar z\,\bigl\{
(G_{\mu\nu}-B_{\mu\nu})\partial X^\mu\bar\partial X^\nu-\half G_{\mu\nu}\bar\psi^\mu\dot\psi^\nu\nonumber\\
&&\ \ \ \ \ \ \ \ \ \ \ \ \ \ -\half\Gamma_{\mu\sigma\rho}\bar\psi^\mu\dot X^\sigma\psi^\rho
-{1\over8}R_{\mu\sigma\rho\tau}\bar\psi^\mu\psi^\rho\bar\psi^\sigma\psi^\tau
\bigr\}.\label{Igfnob}
\eeqa
This corresponds precisely to Witten's topological sigma model\footnote{Our 
variables $\psi, \bar\psi$ and $X$ are identified with the variables $i\chi,-2\psi$ and $\phi$ of \cite{Witten:1991zz}, respectively.} (with a $B$-field 
included).\footnote{We note that the gauge-fixing part of the action can be 
rewritten as  $\{Q_S^+,[Q_S^-,W]\}$ with 
$W\sim\int d^2\sigma G_{i\bar{j}}\bar{\psi}^i \bar{\psi}^{\bar{j}}$,
where $Q_S^+$ and $Q_S^-$ correspond to the $\alpha$ 
and $\tilde{\a}$ part of the transformations in (3.1) of 
\cite{Witten:1991zz}, respectively. Notice that
$Q_S=Q_S^++Q_S^-$. \label{W}}

\subsection{Coupling to topological gravity}

Topological string theory is obtained by coupling the topological sigma model to topological
gravity \cite{Witten:1988xj,Labastida:1988zb}; 
for a review see \cite{Dijkgraaf:1990qw}. The total BRST charge
is the sum of two terms,
\be
Q_{BRST}=Q_S+Q_V,\ \ \ \ \ \ \ \ Q_S^2=Q_V^2=\{Q_S,Q_V\}=0.
\ee
Here $Q_V$ corresponds to the ``usual'' BRST charge $Q_B$ we constructed in section~3 for the
bosonic string, while $Q_S$ corresponds to the charge associated with the symmetry
$\delta_S$ of
the topological sigma model.

As in section~3, the action of $Q_V$ includes
\beqa
\delta_Vg_{ab}&=&2C_\omega g_{ab}+\nabla_ac_b+\nabla_bc_a,\\
\delta_V\tau^k&=&\xi^k,\nonumber \\
\delta_V\tilde b^{ab}&=&\pi^{ab}. \nonumber 
\eeqa
We already know the action of $Q_S$ on the fields of the topological sigma model. Its
action on the fields in the gravitational sector is defined by introducing superpartners for
the fields in the gravitational sector, including the two-dimensional metric $g_{ab}$, the
moduli $\tau^k$, the auxiliary fields $\xi^k$ and $\pi^{ab}$, and the antighost 
field $\tilde b^{ab}$:
\beqa
\delta_Sg_{ab}&=&\psi_{ab},\label{deltaStau} \\
\delta_S\tau^k&=&\hat\tau^k, \nonumber  \\
\delta_S\xi^k&=&\hat\xi^k,\nonumber \\
\delta_Sp^{ab}&=&\pi^{ab},\nonumber \\
\delta_S\beta^{ab}&=&\tilde b^{ab}. \nonumber 
\eeqa
The ghost fields $C_\omega$ and $c^a$ are invariant under $Q_S$.
The $Q_V$ transformation rules of the new fields follow
from the anticommutation relations between $Q_V$ and $Q_S$,
\beqa
\delta_V \psi_{ab} &=&(c^c\partial_c \psi_{ab}+\partial_ac^c \psi_{cb}+\partial_bc^c \psi_{ac}
+2C_\omega \psi_{ab}),\nonumber \\
\delta_V \hat{\t}^k &=& - \hat{\xi}^k, \nonumber \\
\delta_V\beta^{ab}&=&- p^{ab}. \label{betatr}
\eeqa

The topological string worldsheet action is the sum of a 
gravitational term and a sigma model term,
\be \label{topstaction}
L=L_{grav}+L_\sigma.
\ee
The gravitational term is given by
\beqa \label{grfix}
L_{grav}&=&{1\over4\pi} \delta_V\delta_S\left\{\beta^{ab}[g_{ab}-\hat g_{ab}(\tau)]
\right\}\nonumber\\
&=&{1\over4\pi} \delta_V\left\{\tilde b^{ab}[g_{ab}-\hat g_{ab}(\tau)]
+\beta^{ab}[\psi_{ab}-\hat\tau^k\partial_k\hat g_{ab}(\tau)]
\right\}.
\eeqa
and is manifestly invariant under both $Q_V$ and $Q_S$. 
The sigma model term is manifestly invariant under $Q_S$. To ensure invariance
under $Q_V$ we only need to covariantize the gauge fixing fermion in
\eq{BSactiongf}. To do this, we note that $\bar{\psi}^i$ is a $(0,1)$ form 
on $M$ with values in $X^*(T^{1,0})$ and $\bar{\psi}^{\bar{i}}$ is a 
$(1,0)$ form on $M$
with values in $X^*(T^{0,1})$; making explicit the worldsheet index 
we have $\bar{\psi}^i_{\bar{z}}$ and $\bar{\psi}^{\bar{i}}_z$, respectively.
In other words, the worldsheet holomorphic index is correlated with target 
space antiholomorphic index and vice versa. This constraint can be imposed 
covariantly using the projection operator
\be
P^{\m b}_{\n a} = \half (\d^\m_\n \d_a^b - j_a{}^b J^\m{}_\n),
\ee
where $a, b$ are worldsheet indices and $j_a{}^b$ is the worldsheet
complex structure (given by\footnote{With these  conventions, $j^z{}_z=i$
and $j^{\bar{z}}{}_{\bar{z}}=-i$.} 
$j_a{}^b = -\sqrt{g} \e_{a c} g^{c b}$, $\e_{12}=1$). 
We now define
\be
\bar{\psi}'{}^\m_a = P^{\m b}_{\n a} \bar{\psi}^\n_b, \qquad
b'{}^\m_a = P^{\m b}_{\n a} b^\n_b, \qquad
\dot{X}^\m_a = 2 P^{\m b}_{\n a} \partial_b {X}^\n,
\ee
in terms of which the sigma model part of the action is given by
\beqa
\int L_\sigma&=&\int
(\omega_{\mu\nu}+iB_{\mu\nu})d X^\mu\wedge d X^\nu\nonumber\\
&&-{i\over 2}\int d^2\sigma\,\delta_S \left\{\sqrt gg^{ab}
\bar\psi'{}^\mu_a (G_{\mu\nu}\dot X_b ^\nu-{1\over2}G_{\mu\nu}b'{}_b^\nu
+{1\over2}\Gamma_{\mu\sigma\rho}\bar\psi'{}^\sigma_b \psi^\rho)
\right\}.
\label{BSactioncov}
\eeqa
When we work out the action of $\d_S$ in \eq{BSactioncov}, we find
two contributions. The first comes from $\delta_S$ acting on  $\sqrt g g^{ab}$
and on the projection operators; it has the form
\be\label{supercurrent}
\psi_{ab}{\cal G}^{ab}_\s,
\ee
where ${\cal G}_\s$ is the supercurrent of the sigma model. The
second contribution comes from $\delta_S$ acting on the sigma model fields;
it is the familiar sigma model gauge fixing action.

Let us now discuss the insertions in the path integral measure. The terms resulting from 
the $\tilde b$-dependent term in (\ref{grfix}) are the same as in the bosonic string,
so the analysis of the previous section applies. These lead to the usual 
$(\tilde b,\partial_k \hat{g})$ insertions in the path integral. As for the 
remaining terms, integrating out $\psi_{ab}$ sets the auxiliary field $p^{ab}$
equal to the total supercurrent,
\be
p^{ab} = {\cal G}^{ab}.
\ee
Integrating out $\hat{\tau}^k$ then leads to supercurrent insertions 
$({\cal  G},\partial_k \hat{g})$. Finally, integrating out $\hat{\xi}^k$ 
leads to insertions $\delta((\beta, \partial_k \hat{g}))$.

\subsection{Gauge dependence and the holomorphic anomaly}

We would now like to see what happens if we change the gauge for the 
sigma model part of the action. To that end, we consider a change of the 
target space metric, keeping the first line of \eq{BSactioncov} fixed. 
This clearly changes the gauge fixing condition (as follows from the second 
line in (\ref{BSactioncov})). In terms of complexified K\"ahler
moduli, this corresponds to varying the action with respect to the
anti-holomorphic K\"ahler moduli. 
To see this, expand $\omega_{\mu\nu}$ and $B_{\mu\nu}$ in a basis of harmonic
two-forms $\Omega_{I\mu\nu}$,
\be
\omega_{\mu\nu}=\omega^I\Omega_{I\mu\nu},\ \ \ \ \ 
B_{\mu\nu}=B^I\Omega_{I\mu\nu},
\ee
and define the complexified K\"ahler moduli
\be
t^I=\omega^I+iB^I,\ \ \ \ \ \bar t^I=\omega^I-iB^I.
\ee
The original action \eq{BSaction} then depends only on the 
holomorphic moduli $t^I$, while the gauge fixing term in 
\eq{BSactiongf} depends on the combination $t^I+\bar t^I$. 
Varying the metric in the gauge fixing term of the action while keeping the original
action fixed thus corresponds to antiholomorphic deformations. 

It follows that antiholomorphic dependence of correlation functions
is linked to gauge dependence.
In the sigma model there is no such dependence since the
usual argument 
leading to \eq{WI} holds. When coupling to topological gravity, however, 
the correct Ward identity is (\ref{nwi}). The boundary contributions
in (\ref{nwi}) have been computed in \cite{Bershadsky:1993cx}
and do not vanish.%
\footnote{To adapt the BRST Ward identity (\ref{nwi}) to 
the situation in \cite{Bershadsky:1993cx} one should covariantize
the sigma model part as given in footnote \ref{W} and extend 
$Q_S^+$ and $Q_S^-$ to the gravitational sector.}
Thus we conclude that the theory is gauge dependent due to the holomorphic 
anomaly.

{}From the target space point of view, gauge independence of the path 
integral (with respect to these specific deformations of the gauge 
fixing condition) 
would have the interpretation of background independence. The gauge 
dependence due to holomorphic anomaly has been argued to be a manifestation
of a quantum version of background independence \cite{Witten:1993ed}
(see also \cite{Dijkgraaf:2002ac}).
Given that the anomaly originates from boundary terms, it would seem attractive
to try and cancel it via a Fischler-Susskind 
mechanism. In such a scenario the Fischler-Susskind vertex operators 
would shift the background and this could perhaps realize explicitly
the quantum version of background independence. Unfortunately, we have been
unable to find appropriate FS vertex operators.
  
The fact that certain BRST exact states do not decouple due to boundary 
contributions indicates that that there are degrees of freedom localized 
at the boundary of moduli space; these are the would-be gauge degrees of 
freedom that cannot be gauged away because of the anomaly. It would be
interesting to understand the physics associated with these degrees of freedom.
Similar issues arise when one formulates a QFT on a spacetime with 
boundaries. To state two examples, a Chern-Simons theory on a 3-manifold with a
boundary is not gauge invariant (due to the boundary) and induces a 
WZW model at the boundary \cite{Elitzur:1989nr}. 
The second example is AdS gravity: the bulk
diffeomorphisms that induce Weyl transformations on the conformal 
boundary are broken by the holographic Weyl anomaly \cite{HS}. In this case
the trace of the boundary metric cannot be gauged away and in the AdS/CFT 
correspondence acts as a source for the trace of the boundary 
stress energy tensor. The anomalous dependence of the theory on 
a chosen representative of the boundary conformal structure is readily captured
by the anomalous Ward identity similarly to the way the antiholomorphic 
dependence is captured by the holomorphic anomaly.
 
\section{Conclusions}

We have revisited in this paper the BRST quantization of strings.
The main difference with previous treatments is that we use the 
Batalin-Vilkovisky formalism to quantize the worldsheet 
theory. This treatment automatically incorporates the effects 
of zero modes. The extraghosts of the BV formalism are identified 
with the moduli and a measure in moduli space (ghost insertions)
uniquely follows from this
procedure upon integrating out auxiliary fields.
The gauge-fixed action including the ghost insertions is BRST 
invariant. The BRST transformations however
also act on moduli. BRST Ward identities are easily derived
(by adapting the QFT derivation of Ward identities)
and they incorporate terms due to contributions from the boundary 
of moduli space (anomalies). These terms  arise from the fact 
that BRST transformation act by a shift on moduli, so they do not 
leave  the integration domain of moduli invariant. 
We have explicitly discussed bosonic as well as topological strings. In the 
latter example, the BRST Ward identities give rise to the holomorphic
anomaly.

The procedure discussed here is very efficient 
in determining the path integral measure. We should note however 
that additional steps may be required when several patches are needed 
in order to cover the moduli space. In particular, one would need
to pass from local to global data. This may be done by 
utilizing the {\v{C}}ech-De Rham cohomology, as is discussed
in the context of topological gravity in \cite{Becchi}.\footnote{ 
The relevance of the {\v{C}}ech-De Rham cohomology for superstrings is
discussed in \cite{hverlinde} (as cited in \cite{D'Hoker:2002gw, Becchi}).}

It would be  interesting to apply the method discussed here to
quantize the Ramond-Neveu-Schwarz (RNS) superstring. Note that 
a naive integration over moduli space leads to gauge dependent
results at two loops \cite{Verlinde:1987sd}. The  measure at genus~2 
was recently constructed in a series
of paper by D'Hoker and Phong, see \cite{D'Hoker:2002gw} 
for a review. 
It would also be interesting to apply our formalism to derive the 
measure of superstring in the pure spinor formalism \cite{Berkovits:2004px}.
In this respect, we note that it is straightforward \cite{GS} to use the 
methods here to derive the path integral measure for a related model 
\cite{Grassi:2004tv}.

\medskip
\section*{Acknowledgments}
This work was supported by Stichting FOM (B.C.) and by NWO (K.S.). 
B.C.\ thanks the organizers of the program ``String Theory in Curved Backgrounds
and Boundary Conformal Field Theory'' at the Erwin Schr\"odinger Institute in Vienna, of 
the ``Simons Workshop in Mathematics and Physics 2004'' at Stony Brook, and of the 
IPAM ``Conformal Field Theory 2nd Reunion Conference'' at Lake Arrowhead, 
where parts of this work were carried out.




\begin{thebibliography}{mm}


\bibitem{Polchinski:1998rq}
J.~Polchinski,
``String theory. Vol. 1: An introduction to the bosonic string,''
Cambridge, UK: Univ. Pr. (1998).

\bibitem{Deligne:1999qp}
E. D' Hoker, String theory, in 
``Quantum fields and strings: A course for mathematicians.''  Vol. 2, p. 807.
Eds, P.~Deligne {\it et al.}.

\bibitem{D'Hoker:1988ta}
E.~D'Hoker and D.~H.~Phong,
``The Geometry Of String Perturbation Theory,''
Rev.\ Mod.\ Phys.\  {\bf 60}, 917 (1988).


\bibitem{Friedan:1982is}
 D.~Friedan,
``Introduction To Polyakov's String Theory,''
in {\it  Les Houches Summer School, Recent Advances in Field Theory and 
Statistical Mechanics,} ed. Z. Zuber and R. Stora (North-Holland, 1984), p. 839.

\bibitem{Alvarez:1982zi}
 O.~Alvarez,
 ``Theory Of Strings With Boundaries: Fluctuations, Topology, And Quantum
 Geometry,''
 Nucl.\ Phys.\ B {\bf 216} (1983) 125.

\bibitem{Moore:1985ix}
G.~W.~Moore and P.~Nelson,
``Measure for Moduli,''
Nucl.\ Phys.\ B {\bf 266}, 58 (1986).

\bibitem{D'Hoker:1985pj}
E.~D'Hoker and D.~H.~Phong,
``Multiloop Amplitudes For The Bosonic Polyakov String,''
Nucl.\ Phys.\ B {\bf 269} (1986) 205.

\bibitem{Friedan:1985ge}
D.~Friedan, E.~J.~Martinec and S.~H.~Shenker,
``Conformal Invariance, Supersymmetry And String Theory,''
Nucl.\ Phys.\ B {\bf 271} (1986) 93.


\bibitem{Batalin:1984jr}
I.~A.~Batalin and G.~A.~Vilkovisky,
``Quantization Of Gauge Theories With Linearly Dependent Generators,''
Phys.\ Rev.\ D {\bf 28}, 2567 (1983)
[Erratum-ibid.\ D {\bf 30}, 508 (1984)].

\bibitem{Henneaux:1992ig}
M.~Henneaux and C.~Teitelboim,
``Quantization of gauge systems,''
Princeton, USA: University Press (1992).


\bibitem{Baulieu:1987tq}
L.~Baulieu and M.~Bellon,
``BRST Symmetry For Finite Dimensional Invariances: Applications To Global
Zero Modes In String Theory,''
Phys.\ Lett.\ B {\bf 202}, 67 (1988).

\bibitem{Labastida:1988zb}
J.~M.~F.~Labastida, M.~Pernici and E.~Witten,
``Topological Gravity In Two-Dimensions,''
Nucl.\ Phys.\ B {\bf 310} (1988) 611.

\bibitem{Mansfield:1993qd}
P.~Mansfield,
``The Consistency of topological expansions in field theory: 'BRST anomalies'
in strings and Yang-Mills,''
Nucl.\ Phys.\ B {\bf 416}, 205 (1994)
[arXiv:hep-th/9308117].

\bibitem{Becchi}
C.~M.~Becchi and C.~Imbimbo,
``Gribov horizon, contact terms and \v{C}ech- De Rham cohomology in 2D
topological gravity,''
Nucl.\ Phys.\ B {\bf 462}, 571 (1996)
[arXiv:hep-th/9510003].

\bibitem{Cohen:1986mx}
  E.~Cohen, C.~Gomez and P.~Mansfield,
  ``Brs Invariance Of The Interacting Polyakov String,''
  Phys.\ Lett.\ B {\bf 174}, 159 (1986).

\bibitem{Mansfield:1986it}
P.~Mansfield,
``Nilpotent BRST Invariance Of The Interacting Polyakov String,''
Nucl.\ Phys.\ B {\bf 283}, 551 (1987).


\bibitem{Callan:1987px}
C.~G.~.~Callan, C.~Lovelace, C.~R.~Nappi and S.~A.~Yost,
``Adding Holes And Crosscaps To The Superstring,''
Nucl.\ Phys.\ B {\bf 293} (1987) 83.

\bibitem{Polchinski:1987tu}
J.~Polchinski and Y.~Cai,
``Consistency Of Open Superstring Theories,''
Nucl.\ Phys.\ B {\bf 296} (1988) 91.

\bibitem{Bershadsky:1993cx}
M.~Bershadsky, S.~Cecotti, H.~Ooguri and C.~Vafa,
``Kodaira-Spencer theory of gravity and exact results for quantum string
amplitudes,''
Commun.\ Math.\ Phys.\  {\bf 165} (1994) 311
[arXiv:hep-th/9309140].

\bibitem{Witten:1993ed}
E.~Witten,
``Quantum background independence in string theory,''
arXiv:hep-th/9306122.

\bibitem{Dijkgraaf:2002ac}
  R.~Dijkgraaf, E.~Verlinde and M.~Vonk,
  ``On the partition sum of the NS five-brane,''
  arXiv:hep-th/0205281.



\bibitem{Witten:1988xj}
E.~Witten,
``Topological Sigma Models,''
Commun.\ Math.\ Phys.\  {\bf 118}, 411 (1988).

\bibitem{Baulieu:1989rs}
L.~Baulieu and I.~M.~Singer,
``The Topological Sigma Model,''
Commun.\ Math.\ Phys.\  {\bf 125}, 227 (1989).

\bibitem{Witten:1991zz}
E.~Witten,
``Mirror manifolds and topological field theory,''
arXiv:hep-th/9112056.




\bibitem{Dijkgraaf:1990qw}
R.~Dijkgraaf, H.~Verlinde and E.~Verlinde,
``Notes On Topological String Theory And 2-D Quantum Gravity,''
PUPT-1217
{\it Based on lectures given at Spring School on Strings and Quantum Gravity, Trieste, Italy, Apr 24 - May 2, 1990 and at Cargese Workshop on
Random Surfaces, Quantum Gravity and Strings, Cargese, France, May 28 - Jun 1, 1990}.

\bibitem{Elitzur:1989nr}
  S.~Elitzur, G.~W.~Moore, A.~Schwimmer and N.~Seiberg,
  ``Remarks On The Canonical Quantization Of The Chern-Simons-Witten Theory,''
  Nucl.\ Phys.\ B {\bf 326} (1989) 108.

\bibitem{HS}
M.~Henningson and K.~Skenderis,
 ``The holographic Weyl anomaly,''
JHEP {\bf 9807} (1998) 023
[arXiv:hep-th/9806087].

\bibitem{hverlinde} H. Verlinde, ``A note on the integral over fermionic 
supermoduli'', Utrecht preprint No. THU-87/26 (1987), unpublished.

\bibitem{D'Hoker:2002gw}
E.~D'Hoker and D.~H.~Phong,
``Lectures on two-loop superstrings,''
arXiv:hep-th/0211111.


\bibitem{Verlinde:1987sd}
  E.~Verlinde and H.~Verlinde,
  ``Multiloop Calculations In Covariant Superstring Theory,''
  Phys.\ Lett.\ B {\bf 192} (1987) 95.


\bibitem{Berkovits:2004px}
N.~Berkovits,
``Multiloop amplitudes and vanishing theorems using the pure spinor formalism
for the superstring,''
JHEP {\bf 0409}, 047 (2004)
[arXiv:hep-th/0406055].

\bibitem{GS} P.~A.~Grassi and K. Skenderis, unpublished.

\bibitem{Grassi:2004tv}
P.~A.~Grassi and G.~Policastro,
``Super-Chern-Simons theory as superstring theory,''
arXiv:hep-th/0412272.


\end{thebibliography}
\end{document}